\documentclass[10pt,conference]{IEEEtran}
\IEEEoverridecommandlockouts
\usepackage{tcolorbox}
\newcounter{remark}

\usepackage{cite}
\usepackage{amsmath,amssymb,amsfonts}
\usepackage{algorithmic}
\usepackage{graphicx}
\usepackage{textcomp}
\usepackage{booktabs} 
\usepackage{xcolor}
\usepackage{scalerel}
\usepackage{hyperref}
\usepackage{orcidlink}
\hypersetup{
    colorlinks=true,
    linkcolor=black,
    filecolor=black,      
    urlcolor=black,
    citecolor=blue,
    pdftitle={Post-Quantum Cryptography (PQC) Network Instrument: Measuring PQC Adoption Rates and Identifying Migration Pathways},
    pdfpagemode=FullScreen,
    }
\begin{document}

\title{\huge{Post-Quantum Cryptography (PQC) Network Instrument}:\\ Measuring PQC Adoption Rates and \\ Identifying Migration Pathways}

% \author{
% % \IEEEauthorblockN{Phuong Cao\footnote{Corresponding author}}
% % \IEEEauthorblockA{\textit{National Center for Supercomputing Applications} \\
% % \textit{University of Illinois Urbana-Champaign}\\
% % Urbana, IL, U.S. \\
% % pcao3@illinois.edu}
% % \and
% \IEEEauthorblockN{Jakub Sowa}
% \IEEEauthorblockA{\textit{Computer Science} \\
% \textit{University of Illinois Urbana-Champaign}\\
% Urbana, IL, U.S. \\
% jsowa4@illinois.edu}
% \and
% \IEEEauthorblockN{Bach Hoang}
% \IEEEauthorblockA{\textit{Mathematics} \\
% \textit{University of Illinois Urbana-Champaign}\\
% Urbana, IL, U.S. \\
% bachh2@illinois.edu}
% \and
% \IEEEauthorblockN{Steven Qie}
% \IEEEauthorblockA{\textit{Computer Science \& Statistics} \\
% \textit{University of Illinois Urbana-Champaign}\\
% Urbana, IL, U.S. \\
% qie2@illinois.edu}
% \and
% \IEEEauthorblockN{Advaith Yeluru}
% \IEEEauthorblockA{\textit{Computer Engineering} \\
% \textit{University of Illinois Urbana-Champaign}\\
% Urbana, IL, U.S. \\
% ayeluru2@illinois.edu} \\
% \and
% \IEEEauthorblockN{Phuong Cao
% \footnote{Corresponding author}}
% \IEEEauthorblockA{\textit{National Center for Supercomputing Applications} \\
% \textit{University of Illinois Urbana-Champaign}\\
% Urbana, IL, U.S. \\
% pcao3@illinois.edu (*Corresponding author)}
% }

\author{\IEEEauthorblockN{Jakub Sowa \orcidlink{0009-0009-0115-4822}$^1$, Bach Hoang \orcidlink{0000-0003-0827-2564}$^1$, Advaith Yeluru$^1$, Steven Qie$ \orcidlink{0009-0001-4623-131X}^2$,  \\ Anita Nikolich\orcidlink{0000-0003-0407-2805}$^2$, Ravishankar Iyer$^2$, Phuong Cao\thanks{*Corresponding author: Phuong Cao; Data: \href{https://pmcao.github.io/pqc}{https://pmcao.github.io/pqc}} \orcidlink{0000-0001-6028-0583}$^{1,2,*}$
\\}
\IEEEauthorblockA{
\small{
$^1$National Center for Supercomputing Applications,
$^2$University of Illinois at Urbana-Champaign
}}
}

\maketitle

\begin{abstract}
The problem of adopting quantum-resistant cryptographic network protocols or post-quantum cryptography (PQC) is critically important to democratizing quantum computing. The problem is urgent because practical quantum computers will break classical encryption in the next few decades. Past encrypted data has already been collected and can be decrypted in the near future. The main challenges of adopting post-quantum cryptography lie in algorithmic complexity and hardware/software/network implementation. The grand question of how existing cyberinfrastructure will support post-quantum cryptography remains unanswered.

This paper describes: i) the design of a novel Post-Quantum Cryptography (PQC) network instrument placed at the National Center for Supercomputing Applications (NCSA) at the University of Illinois at Urbana-Champaign and a part of the FABRIC testbed; ii) the latest results on PQC adoption rate across a wide spectrum of network protocols (Secure Shell -- SSH, Transport Layer Security -- TLS, etc.); iii) the current state of PQC implementation in key scientific applications (e.g., OpenSSH or SciTokens); iv) the challenges of being quantum-resistant; and v) discussion of potential novel attacks. 

This is the first large-scale measurement of PQC adoption at national-scale supercomputing centers and FABRIC testbeds. Our results show that only OpenSSH and Google Chrome have successfully implemented PQC and achieved an initial adoption rate of 0.029\% (6,044 out of 20,556,816) for OpenSSH connections at NCSA coming from major Internet Service Providers or Autonomous Systems (ASes) such as OARNET, GTT, Google Fiber Webpass (U.S.) and Uppsala Lans Landsting (Sweden), with an overall increasing adoption rate year-over-year for 2023-2024. Our analyses identify pathways to migrate current applications to be quantum-resistant.
\end{abstract}

% \begin{IEEEkeywords}
% \end{IEEEkeywords}

\section{Introduction}

The problem of adopting quantum-resistant cryptographic network protocols or post-quantum cryptography (PQC) is critically important to democratizing quantum computing. The problem is urgent because practical quantum computers~\cite{AnIBMQua39:online}, will break classical encryption in the next few decades. Major applications such as cloud computing~\cite{pham2012toward} (including HPC/supercomputing), financial services, and health analytics~\cite{cao2022predicting} must be migrated to be quantum-resistant. The main challenges of adopting post-quantum cryptography lie in algorithmic complexity and hardware/software/network implementation. The grand question of how existing cyberinfrastructure will support post-quantum cryptography remains unanswered.

This paper describes: i) the \textbf{design of a Post-Quantum Cryptography (PQC) network instrument }placed in a national-scale supercomputing center, ii) the \textbf{latest results on PQC adoption rate} across a wide spectrum of widely used network protocols (Secure Shell -- SSH, Transport Layer Security -- TLS, etc.), iii) the \textbf{current state of PQC implementation }in key scientific applications (e.g., OpenSSH, SciTokens), iv) the \textbf{challenges of being quantum-resistant }and v) discussion of \textbf{potential novel attacks.} Our result is critically important regarding adopting National Institute of Standards and Technology (NIST)'s draft algorithms, such as CRYSTALS-Kyber for encryption, FALCON and SPINCS+ for digital signature, and KEMTALS for key exchange \cite{071pdf62:online} and protocol-specific PQC adaptation such as Hybrid Streamlined NTRU (Ring-Based Public Key Cryptosystem) Prime sntrup761 and x25519 with SHA-512 (sntrup761x25519-sha512) in OpenSSH~\cite{SecureSh29:online,opensshc54:online}.

% This paper describes: i) the design of a Post Quantum Cryptography (PQC) network instrument and the first measurement of PQC adoption in a national-scale supercomputing center across the transportation and application layer of networks (SSH, TLS, RDP) with regard to adopting NIST's being discussed algorithms, such as CRYSTALS-Kyber for encryption, FALCON and SPINCS+ for digital signature, and KEMTALS for key exchange \cite{071pdf62:online} and protocol-specific PQC adaptation such as Hybrid Streamlined NTRU (Ring-Based Public Key Cryptosystem) Prime sntrup761 and x25519 with SHA-512 (sntrup761x25519-sha512) in OpenSSH~\cite{SecureSh29:online,opensshc54:online}.

% Owing to its ancient beginnings, cryptography is one of the most established and well-studied areas in the field of cybersecurity. From the simplest form of encryption, the Caeser cipher, used by various Roman emperors, to modern computers and cellphones connecting to the internet, our methods of hiding, preserving, and authenticating information have always changed with the emergence of new technologies. Cryptography is more important than ever now, though, as hundreds of millions of terabytes of potentially sensitive data are shared daily through public and local networks, including online purchases and banking, health records, legal documents, and even research.

% Thestate69:online

\textbf{Motivation.}
The cryptographic algorithms we use to secure and verify all this data today are far more complex than in centuries past, yet still heavily reliant on the "hardness" of certain mathematical problems. These "hard" problems, such as integer prime factorization and the discrete logarithm problem, form the basis of modern cryptosystems such as RSA and Ecliptic Curve Cryptography (ECC), respectively. As long as no computer can efficiently solve these problems, which are assumed to be hard, the security of the cryptosystem should be maintained. For example, it would take an exascale supercomputer such as Frontier ($\approx$ 1.20 exaflops~\cite{choi2022beating}) longer than the age of the universe, on average, for an attack to reveal the 256-bit private key in the curve25519 ECC cryptographic scheme.

With new developments in quantum computers and quantum algorithms on the rise, many of these problems previously assumed to be hard appear much easier. The 1994 Shor's algorithm for solving both discrete logs and prime factorization efficiently on quantum computers shows the need for this Post-Quantum Cryptography (PQC), which is made even more urgent when you consider an adversary who is storing classically encrypted traffic now to be decrypted with a quantum computer soon when they are powerful and reliable enough. The grand question of how existing cyberinfrastructure will support post-quantum cryptography remains unanswered and will be addressed in this paper.

% \textbf{Challenges.}
% With new developments in quantum computers and quantum algorithms on the rise, many of these problems previously assumed to be hard appear much easier. The 1994 Shor's algorithm for solving both discrete logs and prime factorization efficiently on quantum computers shows the need for this Post-Quantum Cryptography (PQC), is made even more urgent when you consider an adversary who is storing classically-encrypted traffic now, to be decrypted with a quantum computer in the near future when they are powerful and reliable enough. 

% For this reason, cryptographers got to work quickly and began developing numerous quantum-resistant cryptographic algorithms. By 2016, NIST announced a public competition and, in 2022, already chose to standardize some of them: one key exchange scheme, Kyber, and three signature schemes - Dilithium, Falcon, and SPHINCS. In 2024, the standard is expected to be finalized with potentially a few other algorithms, and PQC will start becoming even more widely adopted across networks. 

% In this paper, we study the adoption of PQC on our NCSA network, and suggest potential solutions to the challenges that come with this adoption. More specifically, the key contributions of this work are the following:

\textbf{Key contributions:}
\begin{itemize}
    \item \textit{A novel network instrument to measure PQC adoption in real-time and provides historical trends of the adoption rates across seven layers of computer network protocols.} 
    
    % The instrument is placed at the network border router of nation-scale high-performance supercomputing centers and testbeds.
    
    \item \textit{The first statistical study of PQC's readiness across seven layers of computer network and HPC applications} with a unique vantage point: a petascale supercomputing center with nation-scale visibility through the FABRIC TeraCore network.
    
    \item \textit{Release of an open sample dataset containing metadata of cryptographic suites} across major network protocols (SSH, TLS, RDP, etc. ~\cite{MSRDPBCG15:online}), available online at \href{https://pmcao.github.io/pqc}{\texttt{https://pmcao.github.io/pqc}}
    
    \item \textit{Discussion of potential novel PQC attacks and characterization of major challenges blocking the adoption of PQC} in HPC applications. 
    
    % \item We analyze traffic metadata collected on the NCSA network and determine the extent of adoption of PQC present by identifying cryptographic algorithms already used in network connections. 
    % \item We examine other areas of network traffic metadata that indicate the readiness for new cryptographic implementations. Further, we note ways in which NCSA could improve its readiness for adoption once PQC standards are finalized.
    % \item We propose some general solutions for simplifying network security that could lead to an improved PQC adoption rate including a TLS version 2.0 to distinguish PQC from classical protections as well as TLS termination proxies as a means of consolidating various network protocols' security.
    % \item We discuss areas of future work and tools that could help more easily measure PQC adoption internet-wide. We also discuss the common security issue of downgrade attacks and how it may need further attention specifically with PQC.
\end{itemize}

\textbf{Major results.}
Our results highlight the difficulty of ensuring all systems are updated and using the most secure connection options available, which is important to reach widespread PQC adoptions as follows:
\begin{itemize}

\item OpenSSH and Google Chrome have successfully implemented PQC and achieved an initial adoption rate of 0.029\% (6,044 out of 20,556,816) for OpenSSH connections at NCSA.

\item The adoption rate for OpenSSH is increasing year-over-year for 2023-2024. 

\item Top U.S. large networks (Autonomous Systems -- ASes) such as OARNET, GTT, Google Fiber Webpass, and Comcast; and Uppsala Lans Landsting (Sweden) are hosting clients that have already adopted PQC in OpenSSH. 

\item Over 83\% of Server-side SSH protocols were from 2019 and earlier, 3 years before sntrup761x25519 was even introduced. Only about 65\% of connections used TLS version 1.3, the most recent and most secure option for this use case The rest used TLS version 1.2, which still supports many of the weak cipher suites we found. 

\end{itemize}

\begin{figure}[!b]

\stepcounter{remark}
\begin{tcolorbox}[width=\linewidth, colback=blue!5!white,colframe=blue!75!black,title=Remark \arabic{remark}: \textbf{Our research vs. the state of the art}.]
{
To the best of our knowledge, this is the first network instrument that exists to monitor PQC adoption publicly. There are a few related work~\cite{chen2024gofetch,Thestate69:online,sikeridis2020post,twardokus2022cryptography}, e.g., Cloudflare released a snapshot of TLS adoption in Feb 2024 through their content delivery network~\cite{Thestate69:online}. On the other hand, our network instrument is the first of such to monitor the adoption of both OpenSSH, TLS, and other protocols at large-scale, high-speed (gigabit to terabit) interconnected networks with a wide spectrum of not only traditional cloud but also scientific workloads. 
}
\end{tcolorbox}
\end{figure}

\begin{figure}[!b]

\stepcounter{remark}
\begin{tcolorbox}[width=\linewidth, colback=blue!5!white,colframe=blue!75!black,title=Remark \arabic{remark}: \textbf{Our instrument is tightly integrated with Zeek and network border gateways}.]
{
Our \textbf{Zeek-integrated network instrument to measure PQC adoption} has three main components: 
\begin{itemize}
    \item Parallel and memory-safe log parser to ingest large-volume Zeek logs from a TeraCore network.
    \item Historical analyses scripts that perform regression, trend, and statistics on the parsed metadata of cipher suites.
    \item Real-time snapshot and comparison of our results vs. others (if any) to detect novel attacks and give feedback to NIST and security operators.
\end{itemize}
}
\end{tcolorbox}
\end{figure}

% Our analyses identify pathways to migrate current applications to be quantum-resistant. 

Through experimental deployment at the edges, our approach may help identify empirical novel attacks against PQC implementations and give feedback to the National Institute of Standards and Technology (NIST) and migration pathways for HPC developers~\cite{PostQuan16:online}.

\begin{figure*}
    \centering
    \begin{minipage}{0.93\textwidth}
  \includegraphics[width=\textwidth]{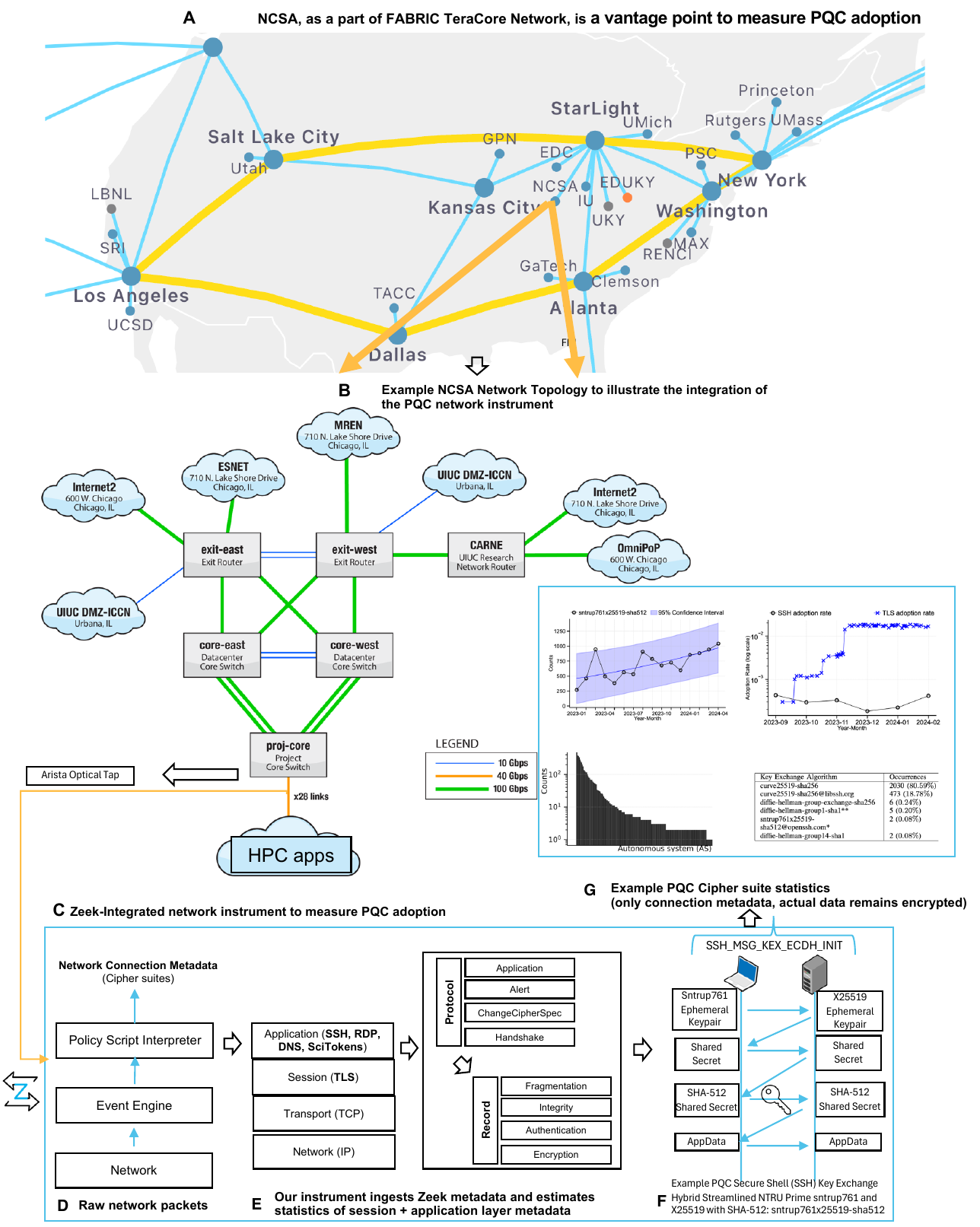}
\centering
\caption{\textbf{Overview of our instrument deployed at NCSA}. Our instrument has visibility into nation-scale network traffic as NCSA is a part of FABRIC testbed (A). We deployed our instrument at the NCSA's network (B) (network topology subject to change) to ingest Zeek connection metadata and parse session and application layer cipher suite information (C,D,E) -- part of C is derived from Zeek documentation and ~\cite{grigorik2013high}. An example of PQC key exchange and the statistical results that we parsed is shown for the Secure Shell (SSH) protocol (F,G).}
\label{fig:overview}
    \end{minipage}%

\end{figure*}

% TODO: \multicolumn{1}{|l|}{\begin{tabular}[c]{@{}l@{}} \textit{Percentage of ciphersuite} \\ \textit{regarding NIST proposal}  \end{tabular}} & TODO         &                                                \\ \hline

\section{Approach Overview}
This section describes the network topology, architecture of the network optical tap, placement of our network instrument, data collected, and statistical methods used. Figure \ref{fig:overview} describes our network instrument. Using FABRIC testbed (A), we deployed our instrument at the NCSA site (B) to ingest Zeek connection metadata and parse session and application layer cipher suite information (C, D, E). An example of PQC key exchange and statistics that we parsed is shown for the Secure Shell (SSH) protocol (F, G). Our instrument is embedded in a nation-scale network for monitoring a wide spectrum of scientific workloads as follows.

\textbf{FABRIC testbed.} FABRIC~\cite{baldin2019fabric} is a nation-scale research infrastructure consisting of a TeraCore network between universities, national labs, and supercomputing facilities such as NSF Cloud testbeds CloudLab, NCSA, and Chameleon (Figure \ref{fig:overview}A). FABRIC powers exploratory networking research at scale in a variety of applications, including cybersecurity, distributed computing, high-speed storage, artificial intelligence, and HPC workloads. The unique approach to FABRIC is its Application Programming Interface (API), allowing experiments to programmatically compose extensible, high-speed interconnected optical link networks, compute, and storage. 

\textbf{NCSA.} The National Center for Supercomputing Applications (NCSA) offers cutting-edge cyberinfrastructure through the National Petascale Computing Facility (NPCF) that houses major NCSA infrastructure, including the Blue Waters supercomputer, to support diverse research needs. Major computing resources include: 1) Delta -- NSF-funded GPU system ideal for GPU-accelerated applications (A100) and gateway-based workflows with nearline storage, Infiniband interconnect; 2) HAL Cluster supporting deep learning system with IBM POWER9 CPUs; and testbeds for ScienceDMZ, honeypot experiments and FPGA boards. All these resources are interconnected with long-term tape archives (Petabytes) and high-speed external connectivity (+400Gbps) to major research networks such as the Energy Sciences Network (ESnet) (Figure \ref{fig:overview}B).

Both NCSA and FABRIC provide a wide gamut of HPC workloads that we will analyze for quantum-resistant cryptography.

\textbf{Zeek network security monitor.}
Zeek (formerly Bro~\cite{paxson1999bro}) is a flexible, open-source platform with decades of development history in network security monitoring. Defenders use Zeek's network analyzers to perform deep packet inspections and provide a broad view of many network protocols. Zeek enables high-performance, high-level, stateful semantic analysis at the application layer and is used operationally at various large sites such as Berkeley Lab and NCSA. The collection and analysis of these data provide a comprehensive view of PQC adoption, as described below.

\textbf{Ethical considerations.}
Our approach works on connection metadata produced by Zeek. We did not use man-in-the-middle or traditional SSL termination equipment. Thus, we only see the encrypted data and not the original content. However, we have access to the timing information of connection establishment to derive the trend and seasonality of the PQC adoption rate. The connection metadata generally includes standard information such as IP address, certificate, and negotiation of cipher suites. We do not see Personally Identifiable Information (PII).

\begin{figure}[!b]

\stepcounter{remark}
\begin{tcolorbox}[width=\linewidth, colback=blue!5!white,colframe=blue!75!black,title=Remark \arabic{remark}: \textbf{Uniqueness of our dataset}.]
{
Our dataset is unique in its placement at a strategic network vantage point at the border gateway router to observe metadata of a wide spectrum of applications across different protocols, validating our PQC measurement for both historical and real-time adoption rates.}
\end{tcolorbox}

\end{figure}

\begin{figure*}[!ht]
    \centering
    \begin{minipage}{.32\textwidth}
        \centering

    \includegraphics[width = \textwidth]{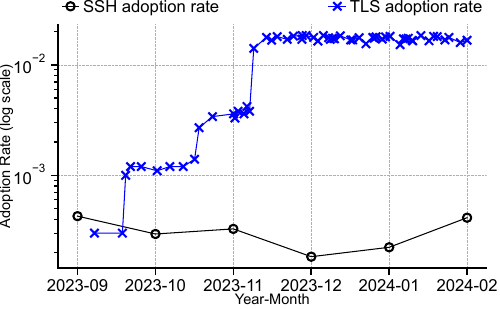}
    \caption{Cross-protocol and cross-site comparison of adoption rate between SSH protocol at NCSA (our analysis) compared with publicly available TLS adoption rate at Cloudflare ~\cite{Thestate69:online}. NCSA records an average of 0.029\% (6044 out of 20,556,816 SSH connections) adoption rate for SSH, while Cloudflare recorded $\approx$ 1.78 percent adoption rate for TLS; more than 99\% adoption came from Chrome~\cite{Thestate69:online}.}
    \label{fig:ssh-pqc}
    \end{minipage}%
\hfill
    \begin{minipage}{.32\textwidth}
        \centering

    \includegraphics[width = \textwidth]{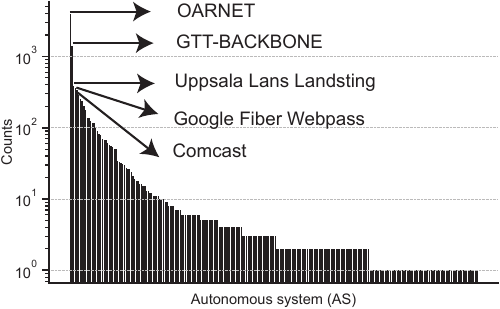}
    \caption{A histogram of autonomous systems adopting PQC in SSH showing that top 5 ASes (OARNE, GTT, Google Fiber, Comcast, etc.)  from U.S. and Uppsala Lans Landsting (Sweden) accounted for the majority of PQC in the head of the distribution. A long list of ASes is shown in the long tail.}
    \label{fig:ssh-as}
    \end{minipage}%
\hfill
    \begin{minipage}{.32\textwidth}
        \centering
    \includegraphics[width = \textwidth]{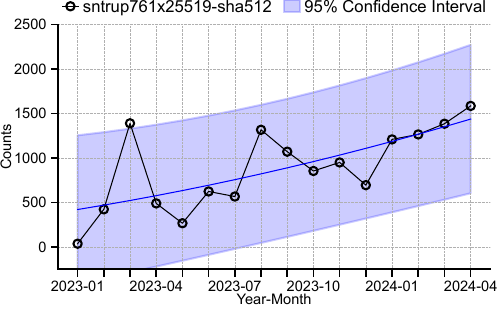}
    \caption{Increasing adoption rate of PQC in SSH key exchange (sntrup761x25519-sha512@openssh.com), starting from January 2023 with only 37 exchanges, up to 1,585 exchanges in April 2024.}
    \label{fig:ssh-keyex}
    \end{minipage}%

\end{figure*}

\begin{table*}[!ht]
\caption{Heterogeneous metadata of cryptographic network protocols powering high-speed scientific workloads at NCSA}
\centering
\begin{tabular}{|p{1.5cm} | p{4cm} | p{10cm}|}
\hline
\multicolumn{1}{|l|}{\textbf{Data Characteristics}}                                               & \textbf{Data Collection}                              & \textbf{Description}                                    \\ \hline
\multicolumn{1}{|l|}{\textit{Number of protocols}}                            & 9 major network protocols &   DNS, Kerberos, Modbus, MySQL, Radius, X509, SSL, SSH, application logs (syslog)                                             \\ \hline
% \multicolumn{1}{|l|}{\textit{Encryption algorithms}}                          & ECDH,  SHA, RSA \cite{10.1145/359340.359342}, etc.                                        &                                                \\ \hline
% \multicolumn{1}{|l|}{\textit{Curve}}                                          & \begin{tabular}[c]{@{}l@{}} x25519,  secp256r1, secp384r1, \\ secp521r1, etc \end{tabular}               &                                                \\ \hline
\multicolumn{1}{|l|}{\textit{Data generator}}                                & Zeek                                                         & Zeek parses raw network packets and produces metadata of network connections.                                               \\ \hline
\multicolumn{1}{|l|}{\textit{Data generation rate}}                           & $\approx$ 30GB compressed logs per day                                                         & Data is compressed in to chunks every hour in the \texttt{gzip} format                                                \\ \hline
\multicolumn{1}{|l|}{\textit{Network speed}}                                  & 400Gbps                                   & The network border links are 400Gbps and is connected to a TeraCore link                                                \\ \hline
\multicolumn{1}{|l|}{\textit{Data amount}}                                      & 13 TB                                                        & Total longitudinal data collected across all seven layers of network                                                \\ \hline
\multicolumn{1}{|l|}{\textit{Format}}                                         & Tab-separated values (tsv)                                                        & Each network protocol has specific fields (source, destination, host key algorithm, etc.)                                              \\ \hline
\multicolumn{1}{|l|}{\textit{Privacy}}                               & Connection metadata                                   &     Only contain metadata of handshake, key exchange, and public certificates (no personally identifiable user data).                                             \\ \hline
\multicolumn{1}{|l|}{\textit{Workload characteristics}}                       & \begin{tabular}[c]{@{}l@{}}Batch, Real-time AI inference, \\ large file transfer (petabytes) \end{tabular} & These workloads make use of the above network protocols, providing a rich source for our analysis.                                                \\ \hline
% \multicolumn{1}{|l|}{\begin{tabular}[c]{@{}l@{}} \textit{PQC ciphersuites} \\ \textit{regarding NIST proposal}  \end{tabular}} & CRYSTAL-KYBER&                                                \\ \hline
\multicolumn{1}{|l|}{\textit{Source and destination}}                         & NCSA and its partner facilities (FABRIC, SDSC, Starlight, ESnet) & Diverse set of partners provide a good vantage point for our analysis.                                                \\ \hline
\multicolumn{1}{|l|}{\textit{Scientific applications}}                        & Representative applications & SciTokens \cite{10.1145/3219104.3219135}, Kubernetes \cite{kubernet9:online}, Kerberos \cite{Kerberos12:online}, Globus \cite{Globus76:online}, and Slurm \cite{SlurmWor96:online} \\ \hline
\multicolumn{1}{|l|}{\textit{Time period}}                                    & 2023-01 to 2024-04 (present)                                                    &      Data are collected in real-time and stored in a network-attached storage system                                          \\ \hline
\multicolumn{1}{|l|}{\textit{Sample PQC protocol}}                                    & Secure Shell (SSH) connection                                                    &      Sample log: \texttt{73.45.xxx.yyy 22 SSH-2.0-OpenSSH\_9.1p1 Debian-2 chacha20-poly1305@openssh.com umac-64-etm@openssh.com sntrup761x25519-sha512@openssh.com ecdsa-sha2-nistp256 }         \\ \hline
\end{tabular}%
\label{tab:data-overview}
\end{table*}

\section{Datasets}
% The dataset is collected from the National Center for Supercomputing Applications (NCSA) by monitoring, identifying, and collecting every network protocol that happened in NCSA $24/7$ between June $1^{st}$ to June $30^{th}$. This data is stored as network logs using the Zeek network analysis framework and collected from a $400$Gbps network border link on the NCSA network. As described in Table \ref{tab:data-overview}, our dataset contains network connections in encrypted protocols such as  Black Hole Router (BHR), Domain Name Service (DNS), Hypertext Transfer Protocol (HTTP), etc. The total size of the dataset is $82$ GB. Since this dataset is collected $24/7$, this dataset can provide the orthogonal view of PQC adoption in scientific network traffic. 

This section presents the dataset that our network instrument has collected, which is built upon our prior work on the security testbed at NCSA~\cite{10.1145/2746194.2746218}. We have collected a heterogeneous dataset containing metadata of cryptographic network protocols powering high-speed scientific workloads at NCSA between 2023-2024. In total, approximately 13TB of metadata has been generated and collected using the Zeek network observability framework from a $400$Gbps network border link on the NCSA network and partner facilities, which supports a wide range of scientific applications \cite{10.1145/3219104.3219135, kubernet9:online, Globus76:online, SlurmWor96:online, Kerberos12:online} such as Globus, SciTokens, and Kubernetes.

A summary of our dataset is specified in Table \ref{tab:data-overview}. Our dataset contains network connections in encrypted application-layer protocols (above layer $3$ TCP/IP), such as Secure Shell (SSH), Remote Desktop Protocol (RDP), Hypertext Transfer Protocol Secure (HTTPS), etc. With metadata including network connections from various sources and destinations, we recorded a wide range of cryptographic algorithms (e.g., ECDH, RSA, SHA-256, etc.) and elliptic curves (e.g., x25519, secp256r1, etc.) that are used for encryptions tasks such as key exchange or key encapsulation mechanism.  Since this dataset is collected $24/7$ for more than 16 months (January 2023 - present), it provides the orthogonal view of PQC adoption in scientific network traffic.

\section{Method \& Results\\ PQC Network Instrument on OpenSSH}
% \textbf{Method and Data} Our method is to collect and analyze metadata of network connections\footnote{E.g., packet headers, handshake algorithms, cryptographic suites, key exchange algorithms, encryption length} in encrypted protocols such as TLS, SSH, and encrypted DNS at a major scientific research backbone National Center for Supercomputing Applications (NCSA) at the University of Illinois. NCSA's 400GBps network border link is the vantage point that major scientific traffic goes through. Measuring metadata at NCSA will provide 360-degree, 24/7, orthogonal view of PQC adoption in scientific network traffic. 

This section presents both: i) longitudinal studies of PQC adoption on the Secure Shell (SSH) protocol over 16 months (Jan 2023 - Apr 2024) and ii) a detailed analysis of cipher suites used in OpenSSH on a sample working day. 

The first and most interesting protocol we studied, owing to its wide usage across most operating systems, was the Secure Shell (SSH) protocol which is well understood through our prior studies~\cite{cao2019caudit,wu2020poster,10.1145/2600176.2600198}. SSH metadata included information on four types of cryptographic algorithms used to secure the connections, the data for each is shown in Table 1. The most immediately apparent detail in the data is that the vast majority of the connections seem to be using some version of OpenSSH. Additionally, there are a few deprecated and insecure cryptographic algorithms in use, which would be a cause for more concern if not for their low usage numbers.

The data, stored as logs using the Zeek network analysis framework, is focused solely on connection metadata. The data was parsed and visualized in Python using its library, Matplotlib, and are shown in Figure~
\ref{fig:ssh-pqc},
\ref{fig:ssh-as},
\ref{fig:ssh-keyex}. We first provide a background on PQC implementation in the latest version of OpenSSH (version $\geq$ 9.0), then describe our results based on SSH data.

\subsection{Background on PQC adoption in OpenSSH}
This section presents details regarding the implementation
of PQ authentication into the OpenSSH library. The first version that OpenSSH supports PQC is 9.0 \cite{opensshc54:online}. 
This developing standard PQC protocol has three main components:
\begin{itemize}
    \item sntrup761 is a key-encapsulation mechanism (KEM) in a family of ring-based public key cryptosystems called Streamlined NTRU Primes  \cite{cryptoeprint:2016/461} \cite{10.1007/BFb0054868}, in which $761$ is a prime number serving as a parameter. Cryptosystems in this family are parameterized by $3$ positive integers $(p, q, w)$, in which $p, q$ are prime numbers, $x^p - x - 1$ is irreducible in the polynomial ring $\mathbb{Z}/q[x]$. The parameters of sntrup761  KEM are a triplet of numbers $p = 761, q = 4591, w = 286$ \cite{ntruprim87:online}.
    
    \item x25519 is the widely implemented Elliptic-Curve-Diffie-Hellman (ECDH) key exchange protocol, using Curve25519 as a underlying curve. Curve25519 \cite{10.1007/11745853_14} is an Montgomery elliptic curve in the form of $y^2 = x^3 +  Ax^2 + x$ over the field $\mathbb{Z}/p$, in which $p = 2^{255} - 19$ is a prime number, and $A = 486662$ is an integer that $A^2 - 4$ is not a square modulo $p$.
    
    \item SHA-512 is a secure cryptographic hash algorithm to ensure data integrity \cite{FederalI41:online}. The input for SHA-$512$ is a message with a size up to $2^{128} - 1$ bits and the output is a word with a consistent length of $512$ bits. 
\end{itemize}

\begin{figure}[!t]
    \centering
  \includegraphics[width=0.4\textwidth]{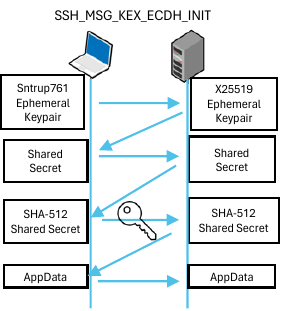}
\centering
\caption{Example flow of Post-Quantum Cryptography Key Exchange Protocol implemented in OpenSSH $\geq$ 9.0 derived from~\cite{SecureSh29:online,opensshc54:online}.}
\label{fig:ssh-pqc-protocol}
\end{figure}

To visualize this PQC protocol, we show key exchange with a client and server in Figure \ref{fig:ssh-pqc-protocol}.

\begin{enumerate}
    \item \textit{Ephemeral Key Generation:}  Client and server generate temporary (ephemeral) key pairs using both sntrup761 and x25519 algorithms.

    \item  \textit{Public Key Exchange:} Client and server exchange their public keys for both sntrup761 and x25519.

    \item   \textit{Shared Secret Calculation:} Each side uses its private key and the other's public key to calculate a shared secret for both the sntrup761 and x25519 algorithms.

    \item \textit{Combining and Hashing:} The two shared secrets are combined and then hashed using the SHA-512 algorithm.

    \item  \textit{Final Key:} The result of this process is the final, strong key used to encrypt and decrypt the SSH communication.
\end{enumerate}

As SSH is widely adopted and frequently updated against vulnerabilities, it is more likely that SSH clients and servers will be the first to adopt PQC schemes. The described protocol is secure because it combines: 1) a hybrid approach, using both a post-quantum resistant algorithm (sntrup761) and a traditional algorithm with strong security (x25519) to provide fault tolerance; and 2) SHA-512, a hashing function securing the final key such that no information about the individual shared secrets can be easily derived.

\begin{figure}[!b]

\stepcounter{remark}
\begin{tcolorbox}[width=\linewidth, colback=blue!5!white,colframe=blue!75!black,title=Remark \arabic{remark}: \textbf{Quantum-resistant property of Streamlined NTRU Prime sntrup761.}]
{
The sntrup761 KEM is a lattice-based post-quantum algorithm. Lattice-based algorithms are relatively new in the cryptography world and are thus not as strongly tested against even pre-quantum adversaries as some more traditional schemes. Many PQC implementations such as this one thus combine a more established cryptosystem to ensure immediate security if just one of the involved algorithms is proven weak \cite{josefsson-ntruprime-hybrid-01}.
}
\end{tcolorbox}
\end{figure}
% \textbf{TODO}
% Explain ZEEK collection pipeline and the network collected on at large (Phuong knows this best)
% Describe dataset sizes and timeframes (when was it collected?)
% Describe log file column headers for just one protocol (e.g SSH)
% Explain how some log files contained ciphersuites and some contained specific cryptographic algorithms of either encryption, MAC, host key, or key exchange. Perhaps define the difference between a ciphersuite and each of the types of crypto algorithms
% Bias: yes there exists bias; explain that we're just looking at the network of one supercomputing center for scientific research which may not generalize to other centers or the internet at large. Also perhaps describe the size of NCSA userbase/network and maybe describe the interconnectedness of the network (again, perhaps Phuong can give better insight on this).

% \begin{figure}[t]
%     \centering
%     \includegraphics[width = 0.45\textwidth]{figs/Size (MB) vs. Protocols.pdf}
%     \caption{Caption}
%     \label{fig:enter-label}
% \end{figure}

% \section{Network Data Analysis}

% The data used in this study was captured over a few hour period in the June 2023, through a 400 Gbps networking border link overlayed on the NCSA network and stored as logs using the Zeek network analysis framework. Only connection metadata was used, meaning there was no access to connection content at any time. The data was parsed and visualized in Python using its library, Matplotlib.

\subsection{Results on PQC protocols adoption in SSH}
First, we compare the adoption rate of SSH with the publicly available adoption rate from Cloudflare. Figure \ref{fig:ssh-pqc} shows cross-protocol and cross-site comparison of adoption rate between SSH protocol at NCSA (our analysis) and TLS 1.3 protocol at Cloudflare ~\cite{Thestate69:online}. NCSA records 0.029\% (6044 out of 20,556,816) monthly adoption rate for SSH, while Cloudflare recorded $\approx$ 1.78 percent adoption rate for TLS 1.3; more than 99\% adoption came from Chrome~\cite{Thestate69:online}.

Figure \ref{fig:ssh-as} shows a histogram of autonomous systems adopting PQC in SSH showing that the top 5 ASes (OARNE, GTT, Google Fiber, Comcast, Advanced Communications Technology, etc.)  from U.S. and Uppsala Lans Landsting (Sweden) accounted for the majority of PQC in the head of the distribution. A long list of ASes is shown in the long tail.

Figure \ref{fig:ssh-keyex} shows the increasing adoption rate of PQC in SSH key exchange (sntrup761x25519-sha512@openssh.com), starting from January 2023 with only 37 exchanges, up to 1,585 exchanges in April 2024.

More interestingly, it is clear that there are no PQ algorithms in use except for one, the key exchange algorithm sntrup761x25519 using the SHA512 hash. This algorithm is an implementation of the post-quantum NTRU Prime cryptographic scheme, specifically a hybrid Streamlined NTRU Prime paired with the x25519 ECDH key exchange method to preserve security against classical adversaries in case a vulnerability in NTRU is found. It was introduced in the OpenSSH version 7.9 and added to the list of defaults in version 9.0 in 2022. It was developed by OpenSSH before the third round selections by NIST took place later that year, which actually ended up removing NTRU after the second round in the process. Unfortunately, we could only identify an insignificant amount of connections using this algorithm for key exchange, 0.08\% for a sample day. The overall average adoption rate is  0.029\% (6044 out of 20,556,816) for the period in which we compare SSH adoption rate with Cloudflare's TLS adoption rate (Figure \ref{fig:ssh-pqc}).

\begin{figure}[!b]

\stepcounter{remark}
\begin{tcolorbox}[width=\linewidth, colback=blue!5!white,colframe=blue!75!black,title=Remark \arabic{remark}: \textbf{Secure Shell (SSH) are among first widely adopted PQC implementation.}]
{
SSH is among the first protocols to implement PQC, albeit with a low adoption rate of 0.029\% recorded at NCSA. However, as our results show, clients increasingly adopt PQC in SSH.
}
\end{tcolorbox}
\end{figure}

\subsection{Results on classical protocols in SSH}
Here, we provide further detailed traffic analysis for a regular working day in June 2023 for the OpenSSH protocol at NCSA. Network metadata was collected from a 400 Gbps network border link on the NCSA network. 

Importantly, we see the presence of various deprecated algorithms used by some connections as shown in Table ~\ref{tab:outdated-ssh}. This poses a security risk outside the scope of PQC and warrants another look at the versions of SSH protocols in use. 

Through further inspection of metadata logs, we determined that over 83\% of Server-side SSH protocols were from 2019 and earlier, three years before sntrup761x25519 was even introduced. This highlights a common problem in the technology space—the difficulty of ensuring all systems are updated and using the most secure connection options available. This is important if we are to ever reach a complete PQC adoption.

\begin{table}[!bpt]
\centering
\caption{Cryptographic algorithms found in sample SSH connection data (*post-quantum, **classical deprecated algorithms as determined by IETF/NIST).}
\begin{tabular}{|p{4.5cm}|p{3cm}|}
\hline
Encryption Algorithm & Occurrences \\\hline
aes256-gcm@openssh.com & 1686 ($66.93\%$) \\
aes128-ctr & 454 ($18.02\%$) \\
chacha20-poly1305@openssh.com & 188 ($7.46\%$) \\
aes128-gcm@openssh.com & 156 ($6.19\%$) \\
aes256-ctr & 31 ($1.23\%$) \\
aes128-cbc & 2 ($0.08\%$) \\
3des-cbc** & 1 ($0.04\%$) \\
\hline
\end{tabular}

\vspace{10 pt}

\begin{tabular}{|p{4.5cm}|p{3cm}|}
\hline
MAC Algorithm & Occurrences \\\hline
hmac-sha2-256-etm@openssh.com & 1844 ($73.20\%$) \\
hmac-sha2-256 & 457 ($18.14\%$) \\
umac-128-etm@openssh.com & 154 ($6.11\%$) \\
umac-64-etm@openssh.com & 33 ($1.31\%$) \\
hmac-sha1 & 17 ($0.67\%$) \\
hmac-sha2-512 & 13 ($0.52\%$) \\
\hline
\end{tabular}

\vspace{10 pt}

\begin{tabular}{|p{4.5cm}|p{3cm}|}
\hline
Host Key Algorithm & Occurrences \\\hline
ecdsa-sha2-nistp256 & 1275 ($50.62\%$) \\
ssh-ed25519 & 1233 ($48.95\%$) \\
ssh-rsa** & 5 ($0.20\%$) \\
rsa-sha2-512 & 4 ($0.16\%$) \\
\hline
\end{tabular}

\vspace{10 pt}

\label{tab:outdated-ssh}
\begin{tabular}{|p{4.5cm}|p{3cm}|}
\hline
Key Exchange Algorithm & Occurrences \\\hline
curve25519-sha256 & 2030 ($80.59\%$) \\
curve25519-sha256@libssh.org & 473 ($18.78\%$) \\
diffie-hellman-group-exchange-sha256 & 6 ($0.24\%$) \\
diffie-hellman-group1-sha1** & 5 ($0.20\%$) \\
sntrup761x25519-sha512@openssh.com* & 2 ($0.08\%$) \\
diffie-hellman-group14-sha1 & 2 ($0.08\%$) \\
\hline
\end{tabular}

% Deprecated KEX:
% https://www.ietf.org/archive/id/draft-ietf-curdle-ssh-kex-sha2-13.html
\end{table}

\section{Method and Results \\ PQC Network Instrument on Other Protocols}
This section describes our analysis extended to other major network protocols such as Remote Desktop Protocol (RDP), Domain Name System (DNS), and Transport Layer Security (TLS). \textit{Note that this analysis is focused on a sample working day to illustrate the adoption.} The objective is to measure the readiness level of applications using these protocols with regard to Post-Quantum Cryptography. As we already provided the trend and seasonality of SSH adoption over the last year, in this part, we will provide a snapshot of other protocols in a working day of collected network traffic.

\subsection{RDP}
The next network protocol we looked at was the Remote Desktop Protocol (RDP) for remote desktop visualization of Windows machines. Although the protocol today has no simple solution for using PQC, it is still worth analyzing the cryptography available given the protocol's prevalence in the IT space and its history as a tool for attackers \cite{sophos}. RDP plays a crucial role for many organizations as a way to manage servers, troubleshooting, and remote work. With the rise in popularity of remote access solutions, ensuring RDP meets future quantum-safe standards is necessary to not only protect the integrity of data transmitted via RDP but also ensure that RDP sessions remain secure in the face of quantum computing advancements.  

Even though for RDP, there was a minimal data set of 26 connections, we were able to identify an interesting issue. Generally, RDP can be configured to use two types of cryptographic security, Enhanced Encryption and Network-Layer Authentication (NLA). Enhanced Encryption for RDP offers encryption over TLS of everything after the Connection Initiation stage and also allows authentication of the server from the client side. NLA requires the client to be authenticated before establishing a session with the server. However, out of all 26 connections, only two of them used the HYBRID-EX security protocol setting, which uses both Enhanced Encryption and NLA. As with SSH, this calls for more careful management of individual machines and making sure they are all configured to use both Enhanced Encryption and NLA for optimal security.

A final thing of note regarding RDP is that on connections where both server and client are using Windows 11 machines, it can be configured to run over TLS 1.3 to add additional levels of security.

\begin{figure}[!b]

\stepcounter{remark}
\begin{tcolorbox}[width=\linewidth, colback=blue!5!white,colframe=blue!75!black,title=Remark \arabic{remark}: \textbf{Initial adoption of TLS v1.3 with hybrid PQC protocol is relatively slow.}]
{
This difficulty in adopting TLS v1.3, though, is still prevalent throughout the internet and not just here at NCSA. It brings to light the unfortunate truth that if widely adopting the newest TLS has taken this many years, PQC adoption may show the same challenge.
}
\end{tcolorbox}
\end{figure}

\subsection{DNS}

We then briefly examined the Domain Name System (DNS) protocol used for deriving IP addresses from domain names. Historically with DNS, packets are not encrypted - this is also true for the NCSA network. However, there are various ways of encrypting DNS traffic, namely through the HTTPS protocol, or over a TLS connection.

HTTPS can be used with DNS, called DoH, to encrypt DNS queries with HTTPS over port 443. This type of DNS can be enabled on some browsers such as Firefox and Chrome. As a side effect, DNS-over-HTTPS makes distinguishing it from general web traffic in logs difficult. 

Another form of secure DNS is DNS-over-TLS (DoT). DoT creates encrypted TLS channels for DNS queries with a DNS-over-TLS domain name resolver instead of the typical unencrypted DNS resolver.

With both these methods of securing DNS and many others not discussed here, it should be noted that any security properties of DNS are transferred from other network protocols, hence creating an opportune security dependence chain discussed further in later sections.

\begin{table}[tbp!]
\centering
\caption{A list of the top 10 cipher suites found in sample TLS connection data (*in TLSv1.3, **considered non-secure).}
\begin{tabular}{|p{4.5cm}|p{3cm}|}
\hline
TLS Ciphersuites & Occurrences \\ &  \\\hline
TLS-AES-128-GCM-SHA256* & 416447 ($53.02\%$) \\
\hline
TLS-ECDHE-RSA-WITH-AES-256-GCM-SHA384 & 117788 ($15.00\%$) \\
\hline
TLS-AES-256-GCM-SHA384* & 100708 ($12.82\%$) \\
\hline
TLS-ECDHE-RSA-WITH-AES-128-GCM-SHA256 & 79171 ($10.08\%$) \\
\hline
TLS-DH-ANON-WITH-AES-256-GCM-SHA384** & 42261 ($5.38\%$) \\
\hline
TLS-ECDH-ANON-WITH-AES-256-CBC-SHA** & 14787 ($1.88\%$) \\
\hline
TLS-ECDHE-RSA-WITH-NULL-SHA** & 5612 ($0.71\%$) \\
\hline
TLS-ECDHE-ECDSA-WITH-AES-128-GCM-SHA256 & 3382 ($0.43\%$) \\
\hline
TLS-ECDHE-RSA-WITH-CHACHA20-POLY1305-SHA256 & 2787 ($0.35\%$) \\
\hline
TLS-ECDHE-ECDSA-WITH-AES-256-GCM-SHA384 & 2497 ($0.32\%$) \\
\hline
\end{tabular}

\end{table}

% \subsection{HTTPS}

\subsection{TLS}

The final network protocol investigated in our study was the Transport Layer Security (TLS) protocol. As with RDP and DNS, there has not been tremendous work finished in the community regarding PQC implementation in this protocol. Some companies and organizations have begun developing specifically tailored PQC implementations for their own use, but little has been developed for the common internet user. Microsoft, for example, remains in the process of collaborating with OpenSSL to integrate PQC into their open-source TLS libraries, but that is still in the works \cite{microsoft-tls-openssl}. With that in mind, it remains to be said that no PQC was found in the TLS network traffic data at NCSA. As with RDP though, it is again worth looking at the security and PQC readiness of TLS here anyway.

Primarily, the security of TLS lies in its various cipher suites, or groupings of cryptographic algorithms, used for securing internet traffic. This data was collected in Table 2, where we show the top ten cipher suites appearing in the connections. While the most commonly found cipher suites in the data are today considered to be secure against classical adversaries, such as the top cipher suite TLS-AES-128-GCM-SHA256, a sizable portion of connections (over 8\%) used weak cipher suites that have a concerning vulnerability.

Even though Chrome browser 116 and above offers TLS 1.3 hybridized Kyber support for PQC, the client does not turn it on by default. In addition, this problem is in part due to the other data point we examined, the TLS versions in use on the server end of the connections: only about 65\% of connections used TLS version 1.3, the most recent and most secure option for this use case. The rest used TLS version 1.2, which still supports many of the weak cipher suites we found. Continuing this message of heightened security maintenance would greatly improve the resilience of network traffic to adversaries at NCSA if all machines were to adopt support for TLS v1.3.

\begin{table*}[tbp!]
\centering
\caption{The current state of adoption of scientific applications and protocols regarding post-quantum cryptography. \\ N/A items show in progress or incomplete information to determine PQC readiness.}
\begin{tabular}{|p{1cm} | p{3cm} | p{7cm} | p{5cm}|}
\hline
\textbf{Protocols} & \textbf{Applications/ Libraries} & \textbf{Descriptions}                                        & \textbf{Quantum Resistant Implemetation}                    \\ \hline
BHR                &  ncsa/bhr~\cite{ncsabhrs61:online}                                & Black Hole Router                                            & N/A                                                       \\ \hline
DHCP               &            Internet  Protocol                     & Dynamic Host Configuration Protocol                          & N/A                                                      \\ \hline
DNS                &          Internet Protocol Suite                        & Domain Name Service                                          & N/A                                                       \\ \hline
DPD                &          Internet Key Exchange                        & Dead Peer Detection                                          & N/A                                                     \\ \hline
HTTP               &          Internet Protocol Suite                        & Hypertext Transfer Protocol                                  & Implement through SSL/TLS                                                       \\ \hline
FTP                &    SFTP                              & File Transfer Protocol                                       & Implement through OpenSSH SCP                                                       \\ \hline
Kerberos           &          krb5 \cite{krb5krb590:online},                        & Network Authentication Protocol                              & N/A     \\ \cline{2-2}       &      GSSAPI \cite{RFC1964T71:online}   &    &                       \\ \hline
Modbus             &               Modbus/TCP Security  \cite{MBTCPSec34:online}                 & Client/Server Data Communication Protocol                    & N/A                                                       \\ \hline
MySQL              &          mysql-server \cite{mysqlmys8:online}                        & Relational Database Protocol                                 & N/A                                                       \\ \hline
NTLM               &                                  & New Technology LAN Manager                                   & N/A                                                       \\ \hline
RADIUS             &           FreeRADIUS \cite{FreeRADI37:online}                       & Remote Authentication Dial-In User Service                   & N/A                                                       \\ \hline
RDP                &          FreeRDP \cite{FreeRDP38:online}                        & Remote Desktop Protocol                                      & N/A                                                      \\ \hline
SIP                &          RTP, SRTP                        & Session Initiation Protocol                                  & N/A                                                      \\ \hline
SMB                &          Samba \cite{sambatea63:online}                        & Server Message Block                                         & N/A                                                       \\ \hline
SSH                & openssh                          & Secure Shell                                                 & sntrup761x25519-sha512@openssh.com   \\ \cline{2-2}
                   & libssh                           &                                                              &  key exchange method                                                      \\ \hline
SSL/TLS                & Open Quantum Safe\cite{openquan43:online}                      & Secure Sockets Layer                                         &             KEM (BIKE, CRYSTALS-Kyber), Signature (CRYSTALS-Dilithium)                                                                                                                                         \\ \hline
SMTP               &                                  & Simple Mail Transfer Protocol                                & N/A                                                       \\ \hline
SciTokens          & scitokens                        & Federated Authorization for Distributed Scientific Computing & N/A                                              \\ \hline
\end{tabular}

% For secure transmission that protects the username and password, and encrypts the content, FTP is often secured with SSL/TLS (FTPS) or replaced with SSH File Transfer Protocol (SFTP)

\label{tab:pqc-adoption}
\end{table*}

\section{Current PQC adoption of network protocols and their applications and libraries}

This section describes the current adoption of critical network protocols regarding post-quantum cryptography (PQC). The problem of migrating applications to be quantum resistant is critical to ensure the security of key exchanges (KEX) and the integrity of digital signatures. As there is an increasing concern about future attacks made by quantum algorithms and computers, knowing the current adoption rate in critical protocols is crucial to ensure that these protocols are all secure against all these quantum computers. \textit{To the best of our knowledge, very few, if any, studies have done a systematic survey of PQC adoption across network protocol layers and the wide spectrum of workloads.}

\begin{figure}[!b]

\stepcounter{remark}
\begin{tcolorbox}[width=\linewidth, colback=blue!5!white,colframe=blue!75!black,title=Remark \arabic{remark}: \textbf{Other than SSH and TLS, other network protocols are not ready for PQC.}]
{
Only SSH and SSL have had PQC implementation with real-world adoptions. The rest of the protocols must develop their quantum-resistant cryptographic system or encapsulate their data in TLS 1.3 (hybrid quantum key exchange).
}
\end{tcolorbox}
\end{figure}

Table \ref{tab:pqc-adoption} described the current PQC adoptions of some of the most important protocols in the world. Some protocols are one of many protocols lying in the applications layer of Internet Protocol Suite such as HTTP, DNS, DHCP, SSH, SSL, etc. In addition, there are widely used communications and authentication protocols such as Kerberos, Modbus, MySQL, SIP, and RDP. As described in Table \ref{tab:pqc-adoption}, most critical protocols are not equipped with quantum-resistant algorithms, except for Secure Shell (SSH) and Secure Socket Layer (SSL). SSH protocol, OpenSSH library has implemented PQC for encryption in sntrup761x25519-sha512@openssh.com key exchange method after the release of its 9.0 version. SSL protocol, OpenSSL has another library called oqsprovider \cite{openquan43:online}, which is a provider to standard OpenSSL to implement quantum-safe cryptography for KEM key establishment in TLS1.3. Some quantum-safe algorithms that oqsprovider has implemented include KEM algorithms (BIKE, CRYSTALS-Kyber) and signature algorithms (Falcon, CRYSTALS-Dilithium).

\section{Case study: Challenges of Migrating \\ SciTokens to PQC}

\begin{figure}[!hb]
\centering
\includegraphics[width=0.5\textwidth]{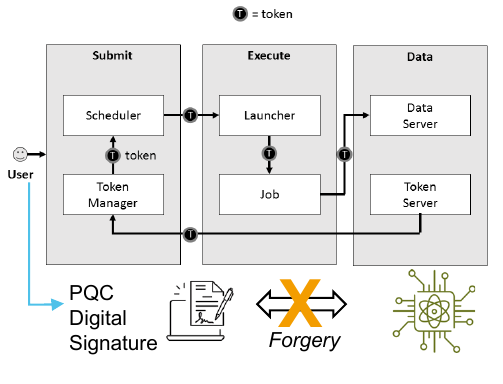}
\centering
\caption{\textbf{Quantum-resistant SciTokens} means the digital signatures in tokens cannot be forged by a Quantum computer. Part of the Figure is adapted from SciTokens illustration~\cite{withers2018scitokens} to illustrate an additional signature scheme needed to be added to SciTokens to be quantum-resistant.}
\label{fig:scitokens}
\end{figure}

SciTokens enables a federated ecosystem for authorization on distributed scientific computing infrastructures, enabling researchers to authenticate themselves as valid users to access scoped scientific computing resources\cite{10.1145/3219104.3219135}. Major supercomputing and scientific instruments (e.g., Laser Interferometer Gravitational-Wave Observatory (LIGO), Open Science Grid, Extreme Science and Engineering Discovery Environment (XSEDE)) rely on SciTokens. It is critical to safeguard SciTokens with quantum-resistant cryptography to avoid the forgery of digital signatures used in tokens (Figure~\ref{fig:scitokens}).

\textbf{Background.}
The SciTokens protocol uses JSON web tokens (JWTs) between various parties, which will validate any access to resources. The process starts off by including information within the JWT. The information included most commonly consists of who issued the token, who the token is assigned to, when the token will expire, and what permissions and access the end user of such token will have. The tokens are then signed cryptographically to be authentic and certified when called upon to use. The benefit of this approach is that each organization or institution can issue its tokens, allowing for easier and quicker access to resources without relying on one central authority to control the distribution of such tokens. 

\textbf{Securing SciTokens to be Quantum Resistant.}
Existing efforts to secure token-based authentications include formal verification~\cite{chen2017svauth}. The next level of security guarantee is to make token-based authentication quantum-resistant. For this, existing software must migrate cryptographic protocols in three main steps: \textit{signing, key exchange, and encryption}. Key exchange and encryption are currently effective, but the looming threat of quantum computing could compromise these safeguards, potentially exposing sensitive data. For SciTokens, the main critical component that must be quantum secure is the \textit{signing} step. When a digital signature is forged, the root of trust about the integrity of the signed token is broken. This undermines trust in historical transaction records, as the digital signatures securing them could be compromised by a quantum computer.

\textbf{Existing work.}
The NIST (National Institute of Standards and Technology) has determined four post-quantum resistant cryptographic algorithms. While there are four determined, one of the algorithms applies to just general encryption, while the other three refer to the digital signing aspect of quantum-resistant cryptography. The general cryptography algorithm is CRYSTALS-Kyber, which is beneficial due to its small encryption keys \cite{NIST:online}. The three digital signing algorithms are CRYSTALS-Dilithium, FALCON, and SPHINCS+, which are distinct from each other due to the high efficiency that CRYSTALS-Dilithium and FALCON provide but also due to the variance in terms of mathematical approach SPHINCS+ has \cite{NIST:online}.

\begin{figure}[!b]

\stepcounter{remark}
\begin{tcolorbox}[width=\linewidth, colback=blue!5!white,colframe=blue!75!black,title=Remark \arabic{remark}: \textbf{Uncertainty in adopting PQC}]
{To the developer community, it is unclear what is the process of determining which post-quantum secure algorithm to use and how to implement those correctly. In the future, we will work with the SciTokens community to correctly implement its quantum-resistant digital signature algorithm.
}
\end{tcolorbox}
\end{figure}

\textbf{Challenges of SciTokens migration to PQC.}

1. As these four algorithms that NIST recommended are relatively new, the mitigation documents are scarce. There are protocol descriptions such as CRYSTALS-Dilithium and Falcon algorithms along with SPHINCS+ \cite{DILITHIUM:online} \cite{FALCON:online} \cite{SPINCS:online}, but implementation guidelines are very few if there are any.

2. Simply adapting SciTokens using existing cipher suites such as RSA-512 will create a new integration problem. For example, extending the length of the RSA digital signature from 256 bits to 512 bits will exceed the length of the HTTP header as defined in RFC 2616, breaking JWT when transmitted through intermediate proxies and interpreted by standard HTTP servers.

3. Emerging side-channel attacks targeting PQC implementations, such as Kyber and Dilithium, show that it is challenging to have a correct implementation, despite that the PQC protocols are correct~\cite{chen2024gofetch}.

\section{Lessons Learned \& \\ Action Items To Preempt Novel PQC Attacks}
\textit{End-to-end PQC requires both client-server support.}
To completely support PQC, we have found that both client and server have to support the same cipher suite. Widespread adoption hinges upon effective protocol negotiation when establishing the initial handshake, followed by key exchanges and digital signatures. For effective post-quantum security to be possible, it is thus necessary that as many machines as possible adopt post-quantum versions of network protocols.

\textit{Internet-wide scan for PQC implementations.}
Another direction is to create a web spider that measures Post-Quantum Cryptography at NCSA and other organizations, constituting a "Network of PQC telescopes" that actively scans IPv4 and IPv6 space for PQC adoption across web services.
In addition, each organization can adopt its own internal PQC compliance checking tool as an HTTPS service so users can visit and perform self-assessments.

\textit{TLS v2.0 supporting PQC by default.}
A more ambitious, yet significant direction for PQC adoption would be the introduction of PQC directly into the TLS standard as TLS v2.0. While some organizations are already working on integrating PQC as additions onto their TLS versions 1.2 and 1.3 \cite{microsoft-tls-openssl}, distinguishing a fully post-quantum implementation with a new version number could prove to be essential in marketing the dire urgency in TLS adoption of PQC internet-wide.

This urgency is highlighted by the incredible potential of securing most network traffic, against both quantum and classical adversaries under TLS. We have shown in earlier sections that there are various methods to secure protocols such as DNS and RDP using just TLS as a wrapper. In addition to those methods, TLS Termination Proxies can also be used as wrappers around current legacy infrastructures to secure traffic seamlessly \cite{f5}. With many TLS-based solutions for cryptographically lacking protocols already out there, it should not be unrealistic to extend this security to the post-quantum world under a new version number, implementation, and standard.

\textit{Observing and Preempting PQC attacks.}
Despite numerous attack attempts ~\cite{cryptoeprint:2024/555} and analyses against current PQC drafts, none of the attacks have been publicly confirmed. We assert that if there were any successful attacks, they would highly likely leave traces in the network metadata, which would be measured by our network instrument or a honeypot~\cite{tay2023taxonomy}, given that we can achieve a widespread deployment on the Internet scale. Therefore, it is important to continue this line of network instruments for PQC measurement to preempt attacks~\cite{pham2014reliability,cao2015preemptive} and failures~\cite{9356415}, particularly in HPC environments. We suspect that futuristic malware~\cite{chung2023stealthml} may employ PQC, in addition to other techniques such as Machine Learning, to hide themselves from forensic analyses.

\textit{PQC downgrade attack.} The security of a PQC scheme depends on both the protocol itself, the implementation, and the effective negotiation of cipher suites (e.g., hybrid) during the migration period. We hypothesize that future attacks may not attempt to exploit the PQC scheme directly but rather downgrade a PQC protocol to a classical protocol, which is less secure, similar to the SSL downgrade attack shown in ~\cite{moller2014poodle}.

\begin{figure}[!t]

\stepcounter{remark}
\begin{tcolorbox}[width=\linewidth, colback=blue!5!white,colframe=blue!75!black,title=Remark \arabic{remark}: \textbf{An open dataset containing metadata of cryptographic suites for major network protocols.}]
{We are releasing an open dataset containing metadata of cryptographic suites across major network protocols (SSH, TLS, RDP, etc.) upon the paper's publication at \texttt{https://pmcao.github.io/pqc}. With our network instrument in place, we will track the adoption of PQC algorithms, give feedback to NIST, expect to discover potential novel PQC attacks in the wild, and help migrate HPC applications to be quantum-safe. Finally, we will expand our network instrument to a set of networked vantage points (PQC telescopes) to continuously measure PQC adoption at the Internet scale.
}
\end{tcolorbox}
\end{figure}

\section*{Conclusion and Future Work}
We have successfully implemented and deployed a novel network instrument to measure PQC adoption in real-time and provide historical trends of the adoption rates across seven layers of computer network protocols at the National Center for Supercomputing Applications at the University of Illinois at Urbana-Champaign. The instrument is placed at the network border router of nation-scale high-performance supercomputing centers and is a testbed for measurements. Using only metadata of network connections, without compromising user's privacy, we present the first statistical study of PQC's readiness across seven layers of computer network and HPC applications with a unique vantage point: a petascale supercomputing center with nation-scale visibility through the FABRIC TeraCore network.

\section*{ACKNOWLEDGEMENTS}
 We acknowledge Dr. Jim Basney for in-depth discussions about SciTokens, Dr. Santiago Nunez-Corrales, Dr. Edoardo Giusto, members of NCSA Quantum Task Force, Dependable Classical-Quantum Computing Systems Engineering Working Group, and the Illinois Quantum Information Science and Technology Center (IQUIST) faculty members, particularly Dr. Brian DeMarco, for insightful feedback. This work was partly supported by the National Science Foundation (NSF) under contract CCF \#2319190. We also want to recognize the following organizations/programs:  the NSF's Trusted CI Cybersecurity Center of Excellence, NCSA Student Pushing Innovation Program (SPIN), Illinois Computes, IBM-Illinois Discovery Accelerator Institute, FABRIC, PPoSS, and the NSF's CyberCorps Scholarship for Service (SFS) Program, ACCESS and Delta allocations, NCSA Integrated Cyberinfrastructure/IRST team, particularly Timothy Boerner, James Eyrich, Ryan Walker, Christopher Clausen, and Dr. Yang Guo at the NIST HPC Security Working Group. Any opinions, findings, conclusions, or recommendations expressed in this material are those of the authors and do not necessarily reflect the views of their employers or the
sponsors.

\bibliographystyle{IEEEtran}
\bibliography{references}

\end{document}